\documentclass[aps,preprintnumbers,twocolumn, superscriptaddress]{revtex4-1}

\usepackage{graphicx}
\usepackage{caption}
\usepackage[shortlabels]{enumitem}
\usepackage{mathtools}
\usepackage{nicefrac}
\usepackage{bbm}
\usepackage{flushend}
\makeatletter

\usepackage{upgreek}

 \usepackage{makecell}
 \usepackage{placeins}
  \usepackage{afterpage}
\usepackage{mathrsfs}

\usepackage{graphics}
\usepackage{amsfonts,amssymb,amsmath}            
\usepackage{ifthen}
\usepackage{amsthm, thm-restate}
\usepackage{dsfont}
\usepackage{float}
\usepackage{MnSymbol}

\usepackage[caption=false]{subfig}
\PassOptionsToPackage{hyphens}{url}
\usepackage[hyperfootnotes=false]{hyperref}
\usepackage{xcolor}
\definecolor{mylinkcolor}{rgb}{0,0,0.7} 
\hypersetup{unicode=true,%
  bookmarksnumbered=false,bookmarksopen=false,bookmarksopenlevel=1, %
  breaklinks=true,pdfborder={0 0 0},colorlinks=true}%
\hypersetup{%
  anchorcolor=mylinkcolor,citecolor=mylinkcolor, %
  filecolor=mylinkcolor,linkcolor=mylinkcolor, %
  menucolor=mylinkcolor,runcolor=mylinkcolor, %
  urlcolor=mylinkcolor} 

\usepackage{tikz}
\usetikzlibrary{arrows}
\usetikzlibrary{arrows.meta}
\usepackage{rotating, xcolor}
\usetikzlibrary{shapes}
\usetikzlibrary{hobby}
\usetikzlibrary{decorations.markings}
\usetikzlibrary{arrows.meta}
\usetikzlibrary{snakes} 
\tikzset{degil/.style={
            decoration={markings,
            mark= at position 0.5 with {
                  \node[transform shape] (tempnode) {$\setminus$};
                  }
              },
              postaction={decorate}
}
}

\newcommand\longrsquigarrow{
\begin{tikzpicture}
\draw [decorate, decoration={zigzag, segment length=+6pt, amplitude=+.95pt,post length=+2pt}, arrows={-Classical TikZ Rightarrow}]  (0,0.1) -- (0.6,0.1); \draw[draw=none] (0,0)--(0.6,0);
\end{tikzpicture}
}

\newcommand\longlsquigarrow{
\begin{tikzpicture}
\draw [decorate, decoration={zigzag, segment length=+6pt, amplitude=+.95pt,post length=+2pt}, arrows={-Classical TikZ Rightarrow},rotate around={180:(0.3,0.1)}]  (0,0.1) -- (0.6,0.1); \draw[draw=none] (0,0)--(0.6,0);
\end{tikzpicture}
}

\makeatletter
\newcommand*{\da@rightarrow}{\mathchar"0\hexnumber@\symAMSa 4B }
\newcommand*{\da@leftarrow}{\mathchar"0\hexnumber@\symAMSa 4C }
\newcommand*{\xdashrightarrow}[2][]{%
  \mathrel{%
    \mathpalette{\da@xarrow{#1}{#2}{}\da@rightarrow{\,}{}}{}%
  }%
}
\newcommand{\xdashleftarrow}[2][]{%
  \mathrel{%
    \mathpalette{\da@xarrow{#1}{#2}\da@leftarrow{}{}{\,}}{}%
  }%
}
\newcommand*{\da@xarrow}[7]{%
  \sbox0{$\ifx#7\scriptstyle\scriptscriptstyle\else\scriptstyle\fi#5#1#6\m@th$}%
  \sbox2{$\ifx#7\scriptstyle\scriptscriptstyle\else\scriptstyle\fi#5#2#6\m@th$}%
  \sbox4{$#7\dabar@\m@th$}%
  \dimen@=\wd0 %
  \ifdim\wd2 >\dimen@
    \dimen@=\wd2 %
  \fi
  \count@=2 %
  \def\da@bars{\dabar@\dabar@}%
  \@whiledim\count@\wd4<\dimen@\do{%
    \advance\count@\@ne
    \expandafter\def\expandafter\da@bars\expandafter{%
      \da@bars
      \dabar@ 
    }%
  }%
  \mathrel{#3}%
  \mathrel{%
    \mathop{\da@bars}\limits
    \ifx\\#1\\%
    \else
      _{\copy0}%
    \fi
    \ifx\\#2\\%
    \else
      ^{\copy2}%
    \fi
  }%
  \mathrel{#4}%
}
\makeatother

\usepackage[nodayofweek]{datetime}
\newcommand{\comment}[1]{}

\newcommand{\obs}{\mathrm{obs}}

\newcommand{\cG}{\mathcal{G}}

\newcommand{\OTP}{\mathrm{OTP}}
\newcommand{\jam}{\mathrm{jam}}
\newcommand{\lp}{\mathrm{loop}}

\theoremstyle{plain}

\theoremstyle{definition}
\newtheorem{definition}{Definition}[section]

\newtheorem{example}{Example}

\begin{document}

\title{Impossibility of superluminal signalling in Minkowski space-time does not rule out causal loops}


\author{V. Vilasini}
\affiliation{Institute for Theoretical Physics, ETH Zurich, 8093 Z\"{u}rich, Switzerland}
\affiliation{Department of Mathematics, University of York, Heslington, York YO10 5DD.}
\author{Roger Colbeck}
\affiliation{Department of Mathematics, University of York, Heslington, York YO10 5DD.}

\date{\today}

\begin{abstract}
Causality is fundamental to science, but it appears in several different forms. One is relativistic causality, which is tied to a space-time structure and forbids signalling outside the future. A second is an operational notion of causation that
considers the flow of information between
physical systems and interventions on them. In [Vilasini and Colbeck, Phys.\ Rev.\ A.\ 106, 032204 (2022)], we propose a framework for characterising when a causal model can coexist with relativistic principles such as no superluminal signalling, while allowing for cyclic and non-classical causal influences and the possibility of causation without signalling. In a theory without superluminal causation, both superluminal signalling and causal loops are not possible in Minkowski space-time. Here we demonstrate that if we only forbid superluminal signalling, superluminal causation remains possible and 
show the mathematical possibility of causal loops that can be embedded in a Minkowski space-time without leading to superluminal signalling. The existence of such loops in the given space-time could in principle be operationally verified using interventions. This establishes that the physical principle of no superluminal signalling is not by itself sufficient to rule out causal loops between Minkowski space-time events. Interestingly, the conditions required to rule out causal loops in a space-time depend on the dimension. Whether such loops are possible in three spatial dimensions remains an important open question.

\end{abstract}

\maketitle

\noindent{\it Introduction.}|Understanding cause-effect relations is central to the scientific method, yet there are several inequivalent notions of causality. Often, it is defined with respect to a background space-time structure, after which causal structure and space-time structure are treated synonymously. An alternative is to define causality operationally and independently of space-time. One way to do this is through causal models, which are based on intervening on physical systems and analysing the resulting correlations~\cite{Pearl2009, Spirtes2001}. This is the approach we take here. Causal models have been extensively applied to situations involving classical variables, being used for instance for medical testing~\cite{Kleinberg2011, Raita2021}, economic predictions~\cite{Spirtes2005, Pearl2009}, 
and machine learning~\cite{Maya2014, Kusumawardani2020, Liu2021}. 

Bell's theorem~\cite{Bell} demonstrates that classical causal models cannot explain quantum correlations within the causal structure that is naturally associated with a Bell experiment~\cite{Wood2015}. This has fuelled several approaches for providing causal explanations to quantum and more general non-classical correlations. One approach is to develop causal models~\cite{Tucci_1995,Leifer_2006,Laskey2007,Leifer_2008,Leifer2013,Henson2014,Wood2015,Pienaar2015,Ried_2015,Costa2016,Fritz_2015,Allen2017,WC_review,Barrett2020A,Pienaar_2020,Schmid2021} that go beyond classical random variables and allow quantum or even post-quantum systems~\cite{Barrett07,WC_causal_GPT} to be causes. Other approaches suggest modifying the causal structure itself without necessarily considering non-classical causes, such as allowing for additional causal influences that go outside the future light cone (e.g., non-local hidden variable theories~\cite{Bohm1952}) or against the direction of time (retro-causality~\cite{Wharton2019}). Such causal influences must remain hidden in order to prevent superluminal signalling at the observed level. More radical approaches lean towards giving up the standard understanding of causation as being acyclic, and replacing it with a suitable notion of logical consistency~\cite{Baumeler_2016, Baumeler2017, Baumeler2018}. While these alternatives correspond to different descriptions of the underlying causal model, they are all compatible with the impossibility of superluminal signalling in Minkowski space-time \cite{Note1}. 

In our associated paper \cite{VilasiniColbeckPRA}, we have developed a general framework for causation that can describe non-classical and cyclic causal influences as well as causation without signalling. We do so by keeping the direction of causation and the order of events in space-time distinct. The former is modelled operationally and we characterize when the two are \emph{compatible} with each other i.e., when we can assign space-time locations to the random variables in the causal model without leading to signalling outside the space-time's future (we call this assignment an embedding).
Here we ask: does the ability to compatibly embed a causal model in an acyclic space-time (such as Minkowski space-time) imply that the operational predictions of the causal model can be reproduced within an acyclic causal structure? If so, it would not be necessary to consider cyclic causation.


Within relativistic physics, causal influences are taken to flow within the light cone, making both causal loops and superluminal signalling impossible in Minkowski space-time. Here we relax this assumption, and require only that observable signalling is limited by the light cone structure. 
In scenarios in which there is causation without correlation (i.e., a \emph{fine-tuned} influence), answering the above question is more challenging since fine-tuned influences can act outside the lightcone without leading to signalling. Our framework allows treatment of both correlations and interventions on physical systems in general scenarios with cyclic causal influences. It can model fine-tuned causal influences as well as latent quantum and post-quantum causes, and be used to characterise conditions under which a causal model can be compatibly embedded in a space-time. Here, we apply this framework to demonstrate the mathematical possibility of causal loops between events in (1+1)-Minkowski space-time where these loops can be operationally detected without superluminal signalling, providing an explicit example. We also show that the observable predictions of this loop cannot be reproduced in any acyclic causal model, answering the aforementioned question in the negative. Interestingly, the same example fails to be embeddable in (3+1)-Minkowski space-time. In the associated paper~\cite{VilasiniColbeckPRA}, we have characterised a large class of operationally detectable causal loops within our framework. Whether any of these may be embeddable in (3+1) dimensions remains an interesting open question.

Fine-tuned causal explanations are often undesirable as fundamental explanations of physical phenomena \cite{Note2}, but can be crucial in practical information processing tasks. For instance, cryptographic protocols (such as the one time pad) rely on fine-tuned correlations, and the security of relativistic cryptographic protocols~\cite{Kent_RBC,CK1} combines both relativistic notions of causality and information-theoretic concepts. Hence, understanding the extent to which compatibility with Minkowski space-time restricts the possible operational causal models also has practical implications. 

\bigskip

\noindent{\it Causal models: correlations, interventions and fine-tuning}|We begin by reviewing the essentials of the causal modelling framework (see~\cite{VilasiniColbeckPRA} for details).
A \emph{causal structure} is modelled as a directed graph $\cG$ whose nodes correspond to observed or unobserved systems and directed edges $\longrsquigarrow$ denote causation between these systems \cite{Note3}. The set of observed nodes (denoted $S_{\obs}$) comprises classical random variables (RVs) such as measurement settings or outcomes, while unobserved nodes may be classical, quantum or post-quantum systems (such as those from a generalised probabilistic theory). 

Implicitly the nodes are associated with \emph{causal mechanisms}  that specify how information from their incoming edges is processed. For instance, in the classical case this processing corresponds to a function $f_{N_i}$ from the parents of the node and possibly an additional, parentless \cite{Note4} RV $E_{N_i}$ (to allow for situations where $N_i$ is not deterministically dependent on its parents) to the node variable $N_i$ itself. ``$A$ is a direct cause of $B$'' then corresponds to the function $f_B$ having $A$ as a (non-trivial) argument. [In the non-classical case, these functions would be replaced by valid maps between systems in the theory (e.g., CPTP maps in quantum theory)].




Often these causal mechanisms are not explicitly known and hence our treatment has to work at the level of observed probability distributions rather than the causal mechanisms. The 
causal structure imposes constraints on the possible distributions that may arise over the observed nodes. One set of such constraints can be expressed using the notion of \emph{d-separation}. For two disjoint subsets $X$ and $Y$ of observed nodes $S_{\obs}$ of a causal structure $\cG$, $X$ and $Y$ are said to be d-separated, denoted $(X\perp^d Y)_{\cG}$, if there are no directed paths between variables in $X$ and $Y$ and if no variables in $X$ and $Y$ have common ancestors in $\cG$. More generally, d-separation $(X\perp^d Y|Z)_{\cG}$ is defined for three disjoint subsets $X$, $Y$ and $Z$ of the observed nodes (see Section~A of the Supplemental Material for details).

We say that a distribution $P_{\cG}(S_{\obs})$ \emph{satisfies the d-separation property} with respect to a causal structure $\mathcal{G}$ if whenever we have d-separation between observed nodes in $\mathcal{G}$, then we have a corresponding conditional independence in the observed distribution $P_{\cG}(S_{\obs})$ i.e., $(X\perp^d Y|Z)_{\cG}$ implies $P_{\cG}(XY|Z)=P_{\cG}(X|Z)P_{\cG}(Y|Z)$. 
If the converse holds for all disjoint subsets of observed nodes, then the distribution is said to be \emph{faithful} or, equivalently, not \emph{fine-tuned}. 
In the present work, we adopt a minimal definition of a causal model which corresponds to a directed graph $\mathcal{G}$ and an observed distribution $P_{\cG}(S_{\obs})$ that satisfies the d-separation property with respect to $\mathcal{G}$ \cite{Note5}.
In this work we allow fine-tuned causal models in which conditional independences can occur in $P_{\cG}(S_{\obs})$ without the corresponding d-separation in $\mathcal{G}$ (i.e., we allow causation without correlation). 

In cases where we know the causal mechanisms, a causal model can be specified by a causal structure $\cG^C$, causal mechanisms $\{f_{N_i}, P_{\cG^C}(E_{N_i})\}_i$ and an associated observed distribution $P_{\cG^C}(S_{\obs})$ for $S_{\obs} \subseteq \{N_i\}_i$. Here the observed distribution must be consistent with relationships specified by the causal mechanisms. Methods for modelling operational causal structures beyond the classical case are proposed in our associated paper~\cite{VilasiniColbeckPRA}. In the present work, we restrict to the classical case which suffices to illustrate our main claims. 



So far, we have only discussed the possible correlations that fit with a causal structure. Inferring causation requires more, and the concept of an intervention has been introduced to deal with this~\cite{Pearl2009}. If intervening on $X$ changes the distribution on $Y$, then we can deduce that $X$ is a cause of $Y$.


In the case where $X$ is parentless in a causal structure $\cG$, correlation between $X$ and another variable $Y$ i.e., $P_{\cG}(Y|X)\neq P_{\cG}(Y)$ suffices to conclude that $X$ is a cause of $Y$. More generally, an intervention on $X$ corresponds to forcing $X$ to take a certain value, $x$, irrespective of its parents. This results in a post-intervention causal structure $\cG_{\text{do}(X)}$ obtained from $\cG$ by removing all the incoming edges to $X$, while maintaining the causal mechanisms for $\cG_{\text{do}(X)}$ from $\cG$, except the mechanism for the intervened node $X$, which is replaced by $X=x$. We say that $X$ \emph{affects} $Y$ if there exist values $x$ and $y$ such that
\begin{equation}
    \label{eq: affects}
    P_{\cG_{\text{do}(X)}}(Y=y|X=x)\neq P_{\cG}(Y=y).
\end{equation} 
In general $P_{\cG_{\text{do}(X)}}(Y|X)$ is not the same as $P_{\cG}(Y|X)$ (but they are equal when $X$ is parentless in $\cG$).




\bigskip

\begin{figure*}[t]
 \centering
\subfloat[\label{fig: OTP}]{\includegraphics[scale=0.8]{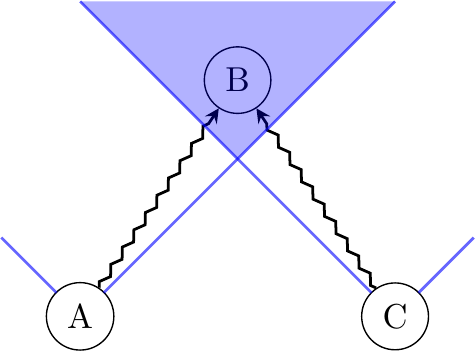}} \subfloat[\label{fig: jamming}]{\includegraphics[scale=0.8]{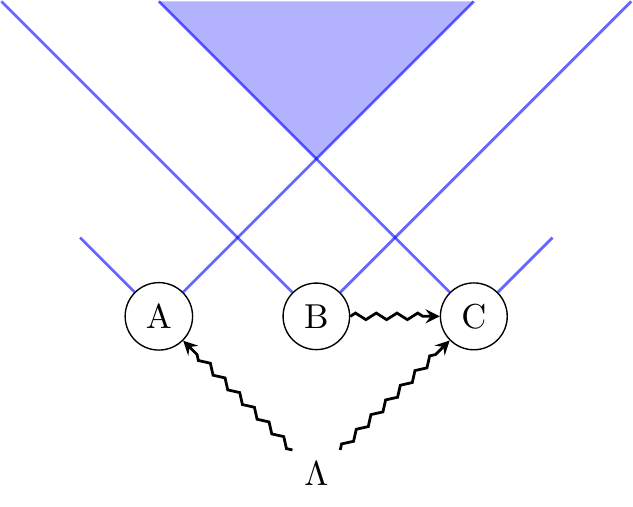}}
\subfloat[\label{fig: loop}]{\includegraphics[scale=0.8]{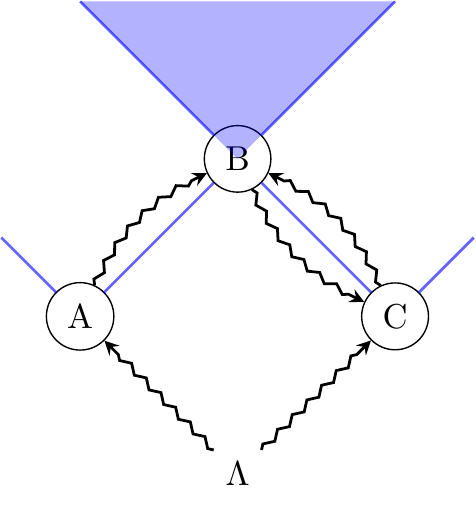}}
  
 \caption{{\bf Three examples of causal models and their compatible embeddings in (1+1)-Minkowski space-time.} 
 In each case, the operational causal structure associated with the model is given in black, circled variables are observed nodes, while uncircled ones are not and the black arrows denote causation. Space-time information is given in blue with time along the vertical and space along the horizontal axis. The solid lines represent light-like surfaces and the shaded region corresponds to the joint future of $A$ and $C$ in all cases. 
 }    
 \label{fig: main}
\end{figure*}

\noindent{\it Acyclic and cyclic causal models embedded in Minkowski space-time}|The affects relation defined in~\eqref{eq: affects} naturally extends to joint interventions on sets of observed nodes. In the presence of fine-tuning, a set of RVs can jointly affect another set without any affects relations between individual pairs of elements in the sets. We will illustrate three such examples here and use these to establish our main claim. 
All three causal models have observed nodes $S_{\obs}:=\{A,B,C\}$ which we will take to be classical and binary. The same observed correlations will be used in all three cases. The distinguishing feature at the observed level will be the affects relations which model what happens under intervention. The causal structures and compatible space-time embeddings for the three examples are given in Figure~\ref{fig: main}. 

\begin{example}[The one-time pad]
\label{example: OTP}
Consider the causal structure of Figure~\ref{fig: OTP}, $\cG^{\OTP}$ with the causal mechanisms $A=E_A$, $C=E_C$, $B=A\oplus C$, with $E_A$ and $E_C$ being independent and uniformly distributed. Then, $B$ is uniformly distributed and we have $P_{\cG^{\OTP}}(B|AC)\neq P_{\cG^{\OTP}}(B)$, $P_{\cG^{\OTP}}(B|A)= P_{\cG^{\OTP}}(B)$ and $P_{\cG^{\OTP}}(B|C)= P_{\cG^{\OTP}}(B)$. Since $A$ and $C$ are parentless, these statements are equivalent to $\{A,C\}$ affects $B$, $A$ does not affect $B$, and $C$ does not affect $B$ \cite{Note6}. A Minkowski space-time embedding of the RVs that does not enable any signalling outside the future is one in which $B$ is assigned a space-time location in the joint future of the space-time locations of $A$ and $C$, as illustrated in Figure~\ref{fig: OTP}. 
\end{example}

\begin{example}[A simplified jamming scenario (cf.~\cite{Grunhaus1996})]
\label{example: jamming}
Consider the causal structure of Figure~\ref{fig: jamming}, $\cG^{\jam}$ with the observed nodes $\{A,B,C\}$ and a classical unobserved node $\Lambda$. Suppose we have the causal mechanisms $\Lambda=E_{\Lambda}$, $A=\Lambda$, $B=E_B$, $C=B\oplus \Lambda$ with $E_{\Lambda}$ and $E_B$ uniformly distributed. This gives the same observed correlations as the previous example, with $B=A\oplus C$ and with all observed variables uniformly distributed. Additionally, $P_{\cG^{\jam}}(AC|B)\neq P_{\cG^{\jam}}(AC)$, $P_{\cG^{\jam}}(A|B)= P_{\cG^{\jam}}(A)$ and $P_{\cG^{\jam}}(C|B)= P_{\cG^{\jam}}(C)$. Since $B$ is parentless, this is equivalent to $B$ affects $\{A,C\}$, $B$ does not affect $A$ and $B$ does not affect $C$. A compatible Minkowski space-time embedding in this case requires that the joint future of the space-time locations of $A$ and $C$ is contained entirely within the space-time future of $B$, as shown in Figure~\ref{fig: jamming}. This is because $B$ affects $\{A,C\}$ can only be verified when $A$ and $C$ are brought together, which is possible only in their joint future. Since we have no pairwise affects relations, there is no pairwise signalling between the RVs and no RV is required to be in the future of any other. Note that the causal influence $B \longrsquigarrow C$ is superluminal, even though there is no superluminal signalling.
\end{example}

\begin{example}[A fine-tuned causal loop]
\label{example: loop}
Consider the causal structure of Figure~\ref{fig: loop}, $\cG^{\lp}$, with the same observed and unobserved nodes as the previous example, but with the causal mechanisms $\Lambda=E_{\Lambda}$, $A=\Lambda$, $C=B\oplus \Lambda$, $B=A\oplus C$ with $E_{\Lambda}$ uniformly distributed.
Note that these causal mechanisms do not admit a unique solution. Nevertheless, in Section~B of the Supplemental Material we apply a method proposed in \cite{VilasiniColbeckPRA} to uniquely determine the observed distribution based on these mechanisms, and show that the same distribution as the previous two examples is obtained.
The effect of interventions in this case cannot be directly inferred from the observed correlations since none of the observed nodes are parentless. The post-intervention causal structure $\cG_{\text{do}(AC)}^{\lp}$ is identical to $\cG^{\OTP}$ and the post-intervention causal structure $\cG_{\text{do}(B)}^{\lp}$ is identical to $\cG^{\jam}$. Applying~\eqref{eq: affects}, we find $\{A,C\}$ affects $B$ \emph{and} $B$ affects $\{A,C\}$ and no pairwise affects relations between $A$, $B$ and $C$. The space-time embedding must satisfy the compatibility conditions of both the previous examples, and Figure~\ref{fig: loop} illustrates an embedding with these properties i.e., this causal loop can be compatibly embedded in (1+1)-Minkowski space-time. 
\end{example}

All three examples above lead to the same correlations $B=A\oplus C$ where $B$ is correlated jointly with $A$ and $C$ but not individually, hence there is fine-tuning. Correlation between $B$ and $\{A,C\}$ implies (by the d-separation property) that $B$ and $\{A,C\}$ must be d-connected i.e., $B$ is d-connected with $A$ and/or $C$. However, the RVs $A$, $B$ and $C$ are pairwise uncorrelated so there must be a pair of variables that are d-connected and yet independent, which constitutes fine-tuning (these independences disappear for small changes in the distribution of one of the variables, e.g., if $\Lambda$ is non-uniform in Examples~\ref{example: jamming} and~\ref{example: loop}). 

The causal loop of Example~\ref{example: loop} exhibits many curious features. It is an operationally detectable causal loop i.e., any causal model that gives rise to the affects relations of this example must necessarily be associated with a cyclic causal structure. This is proven in Section~C of the Supplemental Material, but the intuition is relatively simple: 
Consider three parties in possession of the 3 observed RVs $A$, $B$ and $C$ and two types of experiment: E1) Alice and Charlie perform all possible interventions on $A$ and $C$ and Bob observes $B$ without intervening; E2) Bob performs all possible interventions on $B$ while Alice and Charlie observe $A$ and $C$ without intervening. After both experiments the parties can get together to verify whether $B=A\oplus C$ holds. Here, ``all possible interventions'' on a variable corresponds to running through all possible values of that variable, setting these independently of their parents -- see~\eqref{eq: affects}, and collecting statistics for each choice. These statistics differ between the three causal models, as they have different sets of affects relations. These interventions do not enable the parties to signal outside the space-time future as the affects relations of all these causal models are compatible with the given space-time embedding. 
They nevertheless allow the parties to operationally verify the existence of a causal loop as we show in the Supplemental Material.


In Example~\ref{example: loop}, experiment E1 shows that $A$ and $C$ are causes of $B$ while E2 shows that $B$ is a cause of at least one of $A$ and $C$.  Given the space-time embedding from Figure~\ref{fig: loop}, these interventions would enable the agents to operationally detect retro-causation. 
By contrast, for the first model, E2 would correspond to a post-intervention scenario with no edges and therefore lead to no correlations between the RVs, while in the second model, E1 would also lead to no correlations. In other words, these two experiments enable the parties to operationally distinguish between the causal models of Examples~\ref{example: OTP}-\ref{example: loop} in spite of them having the same observed correlations. 

The mathematical possibility of an operationally detectable causal loop being embedded in Minkowski space-time without signalling necessarily involves fine-tuning (see~\cite{VilasiniColbeckPRA}) 
since in the absence of fine-tuning, signalling and causation are equivalent. Note that in Example~\ref{example: loop} the locations of the random variables in Minkowski space-time have to be carefully chosen to allow compatibility: $B$ and $C$ must be light-like separated, and arbitrarily small adjustments to the location of either $B$ or $C$ will remove compatibility. This requires $B$ to be embedded exactly at the earliest location in the joint future of $A$ and $C$. In a sense this is another kind of fine-tuning, but at the level of the space-time embedding. Furthermore, such an embedding is not possible in (3+1)-Minkowski space-time because the intersection of two lightcones is not itself a cone and consequently there does not exist a frame-independent earliest location in the joint future of two given points, unlike in (1+1)-D~\cite{Note7}.

Beyond these examples, a causal model may give rise to a complicated set of affects relations between various subsets of the observed nodes, making characterising compatibility with a space-time a more complicated task. For instance, if $A$ affects $B$ and $\{A,C\}$ affects $B$ but $C$ does not affect $B$, does $B$ have to be embedded in the joint future of $A$ and $C$, or only in the future of $A$? The novel causal modelling concept required to answer such questions in fine-tuned models is the notion of a \emph{higher-order affects relation}, which we introduce in our framework~\cite{VilasiniColbeckPRA}.  
Here we have illustrated one example, but there is a general class of physical theories involving cyclic causal structures that are compatible with no superluminal signalling in Minkowski space-time as discussed in our associated paper~\cite{VilasiniColbeckPRA}.


\bigskip

\noindent{\it Discussion and outlook.}|In relativistic physics, a causal influence from one space-time location to another implies that the latter is in the future light cone of the former. This means that within relativistic physics, it is not possible to have causal loops or closed timelike curves in space-times whose light cone structures form a partial order. 
On the other hand, the approach adopted here, where causal structure and space-time structure are distinct notions, only requires observable signalling (and not causal influences) to stay within the space-time future for compatibility. 
Maintaining a clear separation between causation and space-time structure and characterising their interdependence is useful for considering formulations of physics without a fixed background space-time structure (e.g., in quantum gravity~\cite{Oreshkov2012, Zych2019}), as well as practical information processing tasks in space-time. 


We note that a previous work~\cite{Horodecki2019} proposes a set of necessary and sufficient conditions for ruling out causal loops in Bell-type scenarios. Our framework~\cite{VilasiniColbeckPRA} identifies implicit assumptions in this claim. In particular, one of the claims in~\cite{Horodecki2019} assumes that no superluminal signalling in Minkowski space-time rules out causal loops, which we have shown does not hold in general.

We have established that the principle of no superluminal signalling in Minkowski space-time alone is insufficient to rule out causal loops, through an explicit construction in (1+1) dimensions. We also found that the conditions for ruling out causal loops can depend on the space-time dimension. Previous works have established that logical consistency~\cite{Deutsch1991, Lloyd2011} or familiar quantum properties such as linearity, no-cloning~\cite{Araujo2015} etc.\ are not sufficient. Finding underlying principles that can do so remains an interesting open problem, a pertinent question being whether the principle of no superluminal signalling rules out causal loops in the case of (3+1)-Minkowski space-time.
\bigskip

\noindent{\emph{Acknowledgements}|} VV acknowledges support from 
the Department of Mathematics, University of York and the ETH Postdoctoral Fellowship from ETH Z\"{u}rich.


\clearpage

\begin{center}
 {\bf SUPPLEMENTAL MATERIAL}   
\end{center}
\section{Further details of the causal modelling framework}
\label{appendix: details}

The d-separation property \cite{Pearl2009, Spirtes2001} was motivated in the main text and can be used as a convenient tool to read off conditional independences in the observed distribution $P_{\cG}(S_{\obs})$ associated with a causal structure $\cG$, from the topology of $\cG$. We provide the formal definition below.

\begin{definition}[Blocked paths]
Let $\mathcal{G}$ be a directed graph in which $X$ and $Y\neq X$ are nodes and let $Z$ be a set of nodes not containing $X$ or $Y$.  A path from $X$ to $Y$ (not necessarily directed) is said to be \emph{blocked} by $Z$ if it contains either $A\longrsquigarrow W\longrsquigarrow B$
with $W\in Z$, $A\longlsquigarrow W\longrsquigarrow B$
with $W\in Z$ or $A\longrsquigarrow W\longlsquigarrow B$ such that neither $W$ nor any descendant of $W$ belongs to $Z$, where $A$ and $B$ are arbitrary nodes in the said path between $X$ and $Y$.
\end{definition}

\begin{definition}[d-separation]
Let $\mathcal{G}$ be a directed graph in which $X$, $Y$ and $Z$ are disjoint
sets of nodes.  $X$ and $Y$ are \emph{d-separated} by $Z$ in
$\mathcal{G}$, denoted as $(X\perp^d Y|Z)_{\cG}$ 
if every path from a variable in $X$ to a variable in $Y$ is \emph{blocked} by $Z$, otherwise, $X$ is said to be \emph{d-connected} with $Y$ given $Z$.
\end{definition}

For acyclic causal structures (both classical \cite{Pearl2009} and non-classical \cite{Henson2014, Barrett2020A}), it is known that any distribution arising from the causal mechanisms satisfies the d-separation property. This is also known to hold in several classical cyclic causal models~\cite{Forre2017} and our accompanying paper~\cite{VilasiniColbeckPRA} provides an example of a quantum cyclic causal model where this holds as well. However, in contrast to the acyclic case, in some classical cyclic causal structures these mechanisms do not imply the d-separation property fails~\cite{Neal2000}. Such examples are ruled out in the definition of causal model we use which imposes the d-separation property.
In the classical case, \cite{Forre2017} introduces a more general graph separation criterion ($\sigma$-separation) that applies to a larger class of cyclic causal models and reduces to d-separation in the acyclic case, and one can in principle generalise our definition of a causal model in terms of $\sigma$-separation to account for this larger class of models. There nevertheless remain classical cyclic causal models for which no graph separation property is known, highlighting the counter-intuitive nature of cyclic causal structures. Characterising the set of classical or quantum cyclic models, defined from causal mechanisms, for which d- or $\sigma$-separation holds is an interesting question for future work, and Appendix C of our accompanying paper~\cite{VilasiniColbeckPRA} outlines a possible method for doing so.

Another useful framework for modelling cyclic quantum causal structures was proposed in \cite{Barrett2020}. This follows a different, split-node approach to causal modelling where all nodes are modelled quantum mechanically. This is a bottom-up approach that describes the observed probabilities in terms of the underlying quantum causal mechanisms.  Unfaithful causal models, post-quantum causes and space-time embeddings of the causal model were not considered here.  In contrast, our framework takes these into account but instead adopts a top down approach where the causal model imposes minimal conditions given by the d-separation property that relates the causal graph with a set of observed distributions. This already suffices to solve the problem regarding space-time embeddings of unfaithful, cyclic and non-classical causal models and reproduces a number of results from the classical causal modelling literature such as Pearl's rules of do-calculus. Even though the causal models in our framework are defined through the d-separation property, the techniques and space-time compatibility condition we propose relies only on the set of affects relations generated by a causal model and can hence also be applied to cyclic causal models that do not satisfy the d-separation condition.

\section{A method to derive the distribution of Example~3}
In acyclic (classical and non-classical) causal models, the observed distribution is uniquely determined once all the causal mechanisms are specified \cite{Pearl2009, Henson2014, Barrett2020A}. However this is not the case in cyclic models. The simplest example illustrating this is the 2-node cycle with $X\longrsquigarrow Y$ and $Y\longrsquigarrow X$ where the causal mechanisms are simply $X=f_X(Y)=Y$ and $Y=f_Y(X)=X$. In this case any distribution $P(XY)=P(X)=P(Y)$ agrees with the causal mechanisms, and using the standard methods available in the previous literature on cyclic causal models \cite{Forre2017, Bongers2021}, this distribution is not uniquely determined by the mechanisms. In the associated paper \cite{VilasiniColbeckPRA} (Appendix C), we have proposed a possible method for uniquely determining the observed distribution in a class of classical and non-classical cyclic models. The method can be applied to models where each directed cycle includes at least one observed node, and involves splitting this observed node in each cycle into two nodes, which results in a directed acyclic graph. We can then use the known methods for acyclic models (which always yield a unique distribution), post-select on the values of the split nodes being equal, and finally renormalise the overall distribution to ensure that the final distribution for the cyclic model is a valid normalised distribution.

We now apply this method to the cyclic causal model of Example~3 from the main text, for which standard methods leave the observed distribution undetermined. The directed cycle in this case is between the nodes $B$ and $C$ and we can choose to split either one of these. Splitting $B$ into two nodes $B$ and $B'$ such that $B$ has all the same incoming arrows as $B$ (and no outgoing arrows) while $B'$ has all the same outgoing arrows as $B$ (and no incoming arrows), we obtain the following acyclic causal model: $P(\Lambda)$ is uniformly distributed, $A=\Lambda$, $C=B'\oplus \Lambda$, $B=A\oplus C$. The corresponding causal structure $\mathcal{G}^{\text{acyc}}$ is given in Figure~\ref{fig:eg3_distributionA}. Notice that the above causal mechanisms for this acyclic graph which are induced by the original cyclic model of Example~3 are not a complete set of mechanisms for the acyclic case, since we are missing a specification of the distribution over the new exogenous node $B'$. Therefore, we can only calculate the conditional distribution $P_{\mathcal{G}^{\text{acyc}}}(ABC|B')$ associated with the observed nodes of the acyclic model. We can then post-select on $B=B'$ and renormalise the distribution to obtain the observed distribution $P_{\mathcal{G}^{\text{loop}}}(ABC)$ of the original cyclic model. This is explicitly shown in Figure~\ref{fig:eg3_distributionB}, and one can see that this procedure yields exactly the same distribution stated in the main text, and in particular this is also the observed distribution obtained in Examples 1 and 2 of the main text. It is easy to verify that splitting the node $C$ instead and applying this procedure also yields the same distribution. Generalising this method to arbitrary non-classical cyclic causal structures is a subject of future work.

\begin{figure}
    \centering
\subfloat[]{\begin{tikzpicture}[scale=0.8]
      \node[shape=circle,draw=black] (B) at (0,0) {$B$};
      \node[shape=circle,draw=black] (B2) at (0,-2) {$B'$};
    \node[shape=circle,draw=black] (A) at (-2,-2) {A};
    \node[shape=circle,draw=black] (C) at (2,-2) {C};  \node[shape=circle,draw=none] (L) at (0,-4) {$\Lambda$};
     \path (A) edge [decorate, decoration={zigzag, segment length=+6pt, amplitude=+.95pt,post length=+4pt}, arrows={-stealth}, thick, bend left=20] (B);
     \path (C) edge [decorate, decoration={zigzag, segment length=+6pt, amplitude=+.95pt,post length=+4pt}, arrows={-stealth}, thick, bend right=20] (B); 

     \path (B2) edge [decorate, decoration={zigzag, segment length=+6pt, amplitude=+.95pt,post length=+4pt}, arrows={-stealth}, thick] (C); 
     \draw[decorate, decoration={zigzag, segment length=+6pt, amplitude=+.95pt,post length=+4pt}, arrows={-stealth}, thick] (L) -- (A);  \draw[decorate, decoration={zigzag, segment length=+6pt, amplitude=+.95pt,post length=+4pt}, arrows={-stealth}, thick] (L) -- (C);
     
      \end{tikzpicture}\label{fig:eg3_distributionA}}\qquad\subfloat[]{\begin{tikzpicture}
 \node[align=center] at (0,0) {\setlength{\extrarowheight}{12pt}\setlength{\tabcolsep}{0.7em}\begin{tabular}{ | c | c | c | c | c| c | c | }
    \hline
   $\Lambda$  & $A$ & $B$ & $B'$ & $C$ & $P_{\mathcal{G}^{\text{acyc}}}(ABC|B')$ & $P_{\mathcal{G}^{\text{loop}}}(ABC)$\\\hline
    0 & 0 & 0 & 0 & 0 & $\frac{1}{2}$ & $\frac{1}{4}$\\ 
    0 & 0 & 1 & 1 & 1 & $\frac{1}{2}$ & $\frac{1}{4}$\\ 
    1 & 1 & 0 & 0 & 1 & $\frac{1}{2}$ & $\frac{1}{4}$\\ 
    1 & 1 & 1 & 1 & 0 & $\frac{1}{2}$&$\frac{1}{4}$\\[4pt]
    \hline
  \end{tabular}};
 \end{tikzpicture}\label{fig:eg3_distributionB}}
    \caption{{\bf Calculating the observed distribution of Example~3} (a) Acyclic causal structure $\mathcal{G}^{\text{acyc}}$ obtained by splitting the node $B$ of the cyclic causal structure $\mathcal{G}^{\text{loop}}$ of Figure~1(b). (b) Table showing the conditional probability distribution $P_{\mathcal{G}^{\text{acyc}}}(ABC|B')$ of the induced causal model over (a). We then post-select on $B=B'$ to recover the original cyclic model and then ignore $B'$. But this yields an unnormalised distribution, and we must therefore renormalise the distribution to obtain the required observed distribution $P_{\mathcal{G}^{\text{loop}}}(ABC)$ as given in the last column.}
    \label{fig:eg3_distribution}
\end{figure}

\section{Operationally certifying the cyclicity of the causal model}
\label{appendix: loop}
We have seen that the cyclic causal model of Example~3 leads to the affects relations $B$ affects $\{A,C\}$, and $\{A,C\}$ affects $B$, with no pairwise affects relations between $A$, $B$ and $C$. We now show that it is impossible for any acyclic causal model to give rise to the same affects relations i.e., these affects relations operationally certify the cyclicity of the causal model and hence the causal loop giving rise to these relations is operationally detectable. 

We show this by contradiction, suppose that there exists an acyclic causal model with an associated causal structure $\cG$, which is a directed acyclic graph (DAG) that produces these affects relations. First consider the affects relation $B$ affects $\{A,C\}$. We will show that this implies the existence of a directed path from $B$ to $A$ or from $B$ to $C$ in $\cG$. Since every acyclic causal model satisfies the d-separation property \cite{Pearl2009, Henson2014}, we can apply it to $\cG$. If there were no directed paths from $B$ to $\{A,C\}$ in $\cG$, then $B$ would be d-separated from $\{A,C\}$ in $\cG_{\text{do}(B)}$ i.e., $(B\perp^d AC)_{\cG_{\text{do}(B)}}$. It is shown in \cite{VilasiniColbeckPRA} that for any $\cG$ satisfying the d-separation property, $(X\perp^d Y)_{\cG_{\text{do}(X)}}$ for any two disjoint sets $X$ and $Y$ of observed nodes implies that $X$ does not affect $Y$. Applying this result here gives $B$ does not affect $\{A,C\}$ which contradicts the given affects relations, so we must have a directed path from $B$ to $\{A,C\}$ in $\cG$.

Next consider a causal model obeying the affects relation $\{A,C\}$ affects $B$ along with $A$ does not affect $B$ and $C$ does not affect $B$. We will now show that this implies the existence of a directed path from $A$ to $B$ and from $C$ to $B$ in $\cG$. For this we will use the concept of a \emph{higher-order affects relation} introduced in \cite{VilasiniColbeckPRA}.  For three disjoint subsets $X$, $Y$ and $Z$ of the observed nodes, we say that $X$ affects $Y$ given do$(Z)$ if there exist values $x$ of $X$, $y$ of $Y$ and $z$ of $Z$ such that
\begin{equation}
    P_{\cG_{\text{do}(XZ)}}(Y=y|X=x,Z=z)\neq P_{\cG_{\text{do}(Z)}}(Y=y|Z=z)
\end{equation}

Note that $\{A,C\}$ affects $B$, and $C$ does not affect $B$ imply the higher-order affects relation $A$ affects $B$ given do$(C)$. Explicitly, the first two conditions are
\begin{align}
    \begin{split}
         P_{\cG_{\text{do}(AC)}}(B|AC)&\neq P_{\cG}(B)\\
          P_{\cG_{\text{do}(C)}}(B|C)&= P_{\cG}(B).
    \end{split}
\end{align}
 When combined, this yields  $P_{\cG_{\text{do}(AC)}}(B|AC)\neq P_{\cG_{\text{do}(C)}}(B|C)$ which precisely expresses the higher-order affects relation $A$ affects $B$ given do$(C)$. Similarly $\{A,C\}$ affects $B$ along with $A$ does not affect $B$ imply the higher-order affects relation $C$ affects $B$ given do$(A)$. It was shown in \cite{VilasiniColbeckPRA} that $X$ affects $Y$ given do$(Z)$ implies that there is a directed path from $X$ to $Y$ in any underlying causal structure $\cG$ producing the higher-order affects relations (and satisfying the d-separation property). Applying this to the two higher-order affects relations obtained above establishes the existence of directed paths from $A$ to $B$ and from $C$ to $B$ in $\mathcal{G}$ (which we have assumed to be a DAG, and this assumption implies that the associated model must satisfy the d-separation property).

To summarise, we have established above that there exists a directed path from $A$ to $B$ and a directed path from $C$ to $B$ in $\cG$ as well as a directed path from $B$ to at least one of $A$ and $C$. This implies the existence of a directed cycle in $\cG$, contradicting the assumption that it is a directed acyclic graph, and certifying the cyclicity of the model.

\end{document}